\documentclass[nofootinbib]{revtex4}
\usepackage[english]{babel}
\usepackage{array,booktabs}   
\usepackage{array} 
\usepackage{amsmath} 
\usepackage{relsize}
\usepackage{wrapfig}
\usepackage{calc}
\usepackage{pdflscape}
\usepackage{color}
\usepackage{float}
\usepackage[font=small,labelfont=bf]{caption}
\usepackage{graphicx}
\usepackage{amsmath}
\usepackage[nodisplayskipstretch]{setspace}

\usepackage{setspace}
\usepackage{tabularx}

\usepackage{float}
\usepackage{color}
\usepackage{amsmath}
\usepackage{float}
\usepackage{calc}
\usepackage{pdflscape}
\usepackage{color}
\usepackage{float}
\usepackage{graphics}
\usepackage{epsfig}
\usepackage{epstopdf}
\usepackage{amssymb}
\usepackage[font=small,labelfont=bf]{caption}
\usepackage{graphicx}
\usepackage{epstopdf}
\usepackage{appendix}
\usepackage{soul}
\usepackage{color}

\definecolor{blizzardblue}{rgb}{0.67, 0.9, 0.93}
\definecolor{bubblegum}{rgb}{0.99, 0.76, 0.8}
\usepackage[urlcolor=blizzardblue]{hyperref}
\hypersetup{
    colorlinks=false,
    linkcolor=blue,
    filecolor=magenta,      
    urlcolor=bubblegum,
}
\begin{document} 

\title{Quantifying multinucleon effect in Argon using high-pressure TPC }

\author{ Jaydip Singh$^{1}$ \footnote{E-mail: jaydip.singh@gmail.com},Srishti Nagu$^{1}$ \footnote{E-mail: srishtinagu19@gmail.com}, Jyotsna Singh$^{1}$ \footnote{E-mail: singh.jyotsnalu@gmail.com}, R.B. Singh$^{1}$ \footnote{Email: rajendrasinghrb@gmail.com}} 

\affiliation{Department Of Physics, University of Lucknow, Lucknow, India.$^{1}$}

\begin{abstract}
Neutrino oscillation experiments use heavy nuclear targets to achieve sufficient interaction rates. Nuclear effects are introduced in the experimental environment by the use of these targets and need to be quantified as they add to the systematic errors. In the low energy region(around 1 GeV) multinucleon events are also present along with Quasi Elastic(QE) and Delta interactions. Therefore if these multinucleon events are not incorporated in the data set properly, we end up with an inaccurate reconstruction of neutrino energy. In our work, we have illustrated the importance of incorporation of multinucleon events for the reduction of systematic errors in physics predictions by DUNE-Near Detector(ND). To achieve this we have presented the event distribution ratio of Ar/C, Ar/Ar, and C/C as a function of squared four-momentum transfer by employing different nuclear models. This analysis recommends the addition of 2p2h or multinucleon events in the event sample and promotes model with Random Phase Approximations(RPA) effect for the analysis of the event sample to overcome or reduce the systematic uncertainties.   
\end{abstract}
\maketitle

\section {INTRODUCTION}	

Precise measurement of neutrino oscillation parameters and understanding of neutrino-nucleus interactions are amongst the current goals to be achieved by the ongoing and future long-baseline neutrino oscillation experiments. The key to unraveling the basics of neutrino oscillation physics is the accurate estimation of the neutrino energy spectrum since the neutrino oscillation probabilities depend on the neutrino energy. Neutrino beams have widely smeared energy owing to the fact that they are produced from secondary decay products. Therefore, the energy of an incoming neutrino must be reconstructed on an event by event basis by observing the properties of the final state. In order to reconstruct the neutrino energy precisely, sufficient knowledge of neutrino-nucleus interactions is apriori. But the environment of the nucleus is such, that it modifies the configuration of the particles that come out of the nucleus along with the kinematics of the interaction. The reconstruction of neutrino energy from the obtained final state particles need to be carefully examined because the particles produced at an initial neutrino-nucleus interaction may not be identical with the final state particles due to the inevitable nuclear effects which are currently in a phase of understanding. Essentially, the identification of nuclear effects that are inherently different for neutrino and antineutrino scattering is important and required to be investigated. This entails a detailed study of (anti)neutrino-nucleus scattering cross-section which gets modified by the nuclear environment of heavy nuclear targets. To collect sufficient statistics, long-baseline neutrino oscillation experiments use massive nuclear targets i.e. Argon(A=40: isovector component), Calcium(A=40: isoscalar target), Carbon(A=12: isoscalar target), Oxygen(A=16: isoscalar target), (where 'A' is the mass number) instead of low mass nuclear targets i.e. Hydrogen(A=1) and Deuterium(A=2). These targets are comparatively cleaner targets as they are devoid of nuclear effects but result in lesser statistics. Thus the data sets obtained by using heavy nuclear targets have reduced statistical uncertainty but are still afflicted by systematic uncertainties. Therefore in a heavier target, we need a handle to curb systematic errors arising due to nuclear effects. Much experimental cross-section data is available for carbon or hydrocarbon CH \cite{ch1,ch2,ch3,ch4,ch5} but very few measurements are there with heavy targets like Argon \cite{ar1,ar2}, Water \cite{wa1,wa2} and Iron \cite{fe1,fe2,fe3}. There is a need to enhance our understanding regarding the nuclear effects present in various targets as this will lead to the development of a better and uniform nuclear model.

In addition to nuclear effects, the few-GeV energy region is complicated because there is an overlap of many reaction channels, physics mechanisms, etc. In this region, a significant contribution arises from the two-body current also known as two-particle-two-hole (2p2h) interaction process, whose experimental signature is 1$\mu^{-}$ and 0$\pi$ which is indistinguishable from a true QE signature. These 2p2h events account for a high cross-section of QE-like events \cite{lalmosgall}. In 2p2h interaction, the incoming neutrino interacts with a correlated pair of nucleons. In this process(energy and momentum transfer to two nucleons) two nucleons are ejected from the nucleus while in the QE interaction process a single recoiling nucleon is ejected. Therefore, multinucleon events are identified by the presence of multiple nucleons in the final state. These multinucleon events have a dominant contribution from 2p2h processes \cite{2p2h}. The 2p-2h events are also known as the Meson Exchange Current(MEC) events since the concerned initial state nucleons keep on exchanging a pion. They are mostly present in the region between QE and $\Delta$ resonance production. Ideally, for a MEC event to occur, the energy transferred to the hadronic system must be more than that for a charged current (CC)QE event. The hadronic re-interactions within the nuclear medium also known as Final State Interactions(FSI) play a crucial role in modifying the initial state products that were actually produced at the initial neutrino-nucleon interaction vertex. The current studies which are related to FSI effects and their inclusion in Monte Carlo(MC) codes can pose difficulty in the measurement of such events. Focussed studies on pion have been accomplished to understand the pion production \cite{pion} data on nuclear targets. In addition to pion events, 2p2h or multinucleon events can also appear as fake QE events and needs to be examined carefully to benchmark FSI effects arising due to multiple nucleons. These uncertainties are also a hurdle in the determination of the CP violating phase which is currently unknown and is one of the most important question to be addressed. Determination of precise CP phase value is not only important for verifying the baryon asymmetry of the Universe \cite{baryon} but also for explaining the effective mass of neutrino \cite{jdsterile}.

Monte Carlo event generators serve as a tool to simulate neutrino scattering interactions and enhance our understanding of the interaction processes involved therein. These event generators are developed by different theoretical communities using different nuclear models. The prediction of the neutrino-nucleus event rates for any particular nuclear target along with the topology of final state particles depends upon the theoretical consideration of the generators. It is critically important that input to these generators must contain the best knowledge of neutrino-nucleus cross sections and nuclear effects. Cross-sectional and nuclear uncertainties occurring in different models, impact significantly the understanding of CP violation, mass hierarchy, and octant sensitivities \cite{srish,srish1,snaaz}. There is an insufficient description of nuclear effects in the Monte Carlo generators presently in use. For eg. the Fermi Gas Model \cite{rfg1, rfg2} which is used to describe the nuclear structure includes fermi motion and Pauli blocking to define nuclear effects in the generators but are inadequate for a complete description of nuclear effects especially for QE and $\Delta$(1232) processes. 

It is of crucial importance for the presently running experiments like T2K \cite{t2k} and NO$\nu$A \cite{nova} and the future long-baseline neutrino oscillation experiments like DUNE \cite{dune2020} and HyperKamiokande \cite{hk} to very well understand neutrino-nucleus interactions. One of the upcoming and the most promising neutrino oscillation experiment, the Deep Underground Neutrino Experiment(DUNE) proposed to be built in the U.S., is designed to deliver precision results related to neutrino oscillation physics. It will consist of a near detector(ND) \cite{duneNDcdr} and a far detector(FD) \cite{dune2020} placed at a distance of 575m and 1300km respectively from a mega-watt facility situated at Fermilab, producing a muon-neutrino dominated beam. The FD will be located 1.5km underground at Sanford Underground Research Facility(SURF), South Dakota, while the ND will be located at Fermilab. The FD will be built up using the liquid Argon TPC(Time Projection Chamber) technology that will provide a detailed view of particle interactions. It will be composed of four modules, 10 kton each, with a total mass of 40 kton and a fiducial mass of 34 kton. The high-pressure gas TPC is one of the considered designs for the ND \cite{duneNDcdr} which uses gaseous Argon as target material. The use of gaseous Argon will enhance the capability of the ND to measure charged particles even with lower energies.  

In this work, we present a simulation-based analysis using GENIE(version-2.12.6) \cite{3} for the DUNE-ND HPgTPC(High Pressurised gas Time Projection Chamber) design to quantify the multi-nucleon contribution in CCQE $\nu(\bar\nu)$-nucleon interactions on Argon and Carbon targets. These targets are extensively used in neutrino detectors, therefore this study will help in the estimation of the uncertainties stemming from the difference between nuclear effects in Argon and Carbon. To incorporate the effects of long range nucleon-nucleon correlations we have selected a model which includes the Random Phase Approximation(RPA) effect in itself. The RPA effect is prominent where the four-momentum transferred to the nucleus during a neutrino-nucleon interaction is small, as the energy transferred to the nucleus increases, the RPA effect diminishes. The detailed study of this effect has been discussed in \cite{multin}.

We present our results as a function of the kinematic variable $Q^{2}$(squared four-momentum transfer) which is sensitive to the presence or absence of multinucleon correlations. To quantify the amount of uncertainty in the measurables due to the use of different targets, the event ratios such as Ar/C, Ar/Ar and C/C are estimated as a function of $Q^{2}$. We have analyzed the QE interaction by using the RPA correlations and adding to it the contribution from 2p2h interactions. The motivation behind this combination is the improvement observed in the description of MiniBooNE data \cite{mini1,mini2,mini3}, such an approach has proved to describe data by previous studies as well, which one can find in \cite{lalmosgall2p2h,rgran}.

The uncertainty in the nuclear models for Carbon target was recently published by the MINER$\nu$A collaboration \cite{main} in which they have shown difference in physics analysis when different QE models are used for the same set of data sample. They performed the analysis by considering the data sample once with and without RPA effect and by adding a 2p2h event sample to the sample with the RPA effect. The study suggests a requirement of modification in the models. More recent results from MINER$\nu$A collaboration \cite{latestMinerva} state that they could match their previous result \cite{main} only on further tuning the 2p2h model using an empirical fit to the hadronic energy spectrum. Their results indicate that different models observe various level of tension with the data and no single model was able to consistently reproduce the data.

The paper is organized in the following sections: In Section II, we describe the currently proposed design of the DUNE-ND. The simulation and experimental details are described in Section III. Section IV includes a description of the multinucleon effect and its implications. The results and the conclusions drawn from our analysis are discussed in Section V and VI respectively. 

\section{DUNE-ND DESIGN: HIGH-PRESSURE GAS TPC }
The concept of a ND is introduced to achieve more detailed information regarding the kinematics of neutrino interactions and it acts as a necessary factor in reducing flux and background signal related uncertainties. An effective way to reduce the systematic uncertainties occurring in the neutrino oscillation experiments is by measuring a non-oscillated event distribution at a ND system and further predicting the oscillated event distribution at the FD. Designing the ND and FD with identical functionalities and targets helps in a maximum cancellation of correlated uncertainties. In order to accomplish DUNE physics goals, a ND will play a crucial role by precisely determining the neutrino energy spectrum, flavor composition and the cross-section of various neutrino scattering processes. The main advantage of the ND will be the collection of a large amount of unoscillated neutrino interaction event rates as it will be exposed to an intense flux of neutrinos. The reference design of the DUNE-ND \cite{hpgtpc3mev, HPgTPC, hpgtpc} consists of three main components: (1) A 50t LArTPC with pixelated readout (2) A multi-purpose tracker, the HPgTPC, kept in a 0.5T magnetic field and surrounded by  ECAL (3) A 8t 3-Dimensional Scintillator Tracker Spectrometer(3-DST). 

The working principle of LArTPC to observe $\nu$-Argon interactions will be similar to that of the FD while the 3-DST will look for neutrino-CH interactions and is designed to have a powerful detection capability for neutrons. The DUNE HPgTPC \cite{hpgtpc3mev} which is a 'copy' of ALICE \cite{alice} TPC will be a single cylindrical volume of diameter 5.2m and length 5m (2.5 m + 2.5 m drift). The gas in HPgTPC is proposed to be a mixture of Argon-$CH_{4}$, with 90$\%$ Argon and 10$\%$ $CH_{4}$(known as P10 gas) and filled at a pressure of 10 Atm. In this mixture 97$\%$ of total interactions will occur via the Argon nuclei. The HPgTPC will have a total volume of $\simeq$ 1.8 ton and a fiducial volume of $\simeq$ 1 ton. It can detect low energy charged particles better than LArTPC due to the use of lower-density active material, which allows the particles to travel a longer path before they finally stop. Charged particles can be tracked over the full 4$\pi$ solid angle within the fiducial volume of the HPgTPC tracker.
It will be advantageous to use HPgTPC in conjunction with LArTPC as both are using the same target material. Due to low energy detection thresholds, HPgTPC will provide better vertex resolution for the identification of low energy particles and will also help in the discrimination of charged particles on an event by event basis. The detection thresholds that we have used in our work are-(1) we apply a momentum cut of 200 MeV \cite{duneNDcdr} for muons, which is equivalent to a total energy cut of 226 MeV. (2) The HPgTPC will be able to detect protons with K.E. $>$ 20MeV with an efficiency of more than 80$\%$, so we have selected a 20 MeV \cite{duneNDcdr} cut on protons. While the minimum K.E. threshold for protons in the HPgTPC can be 3 MeV \cite{hpgtpc3mev}, so this fact can be revised in a future analysis. (3) The K.E. cut for neutrons lies between 50 MeV to 700 MeV \cite{duneNDcdr}. The oscillation studies by DUNE \cite{duneNDref92} have found that the HPgTPC detector is sensitive to 20$\%$ systematic variations in the energy that neutrons may carry away.

\section{SIMULATION AND EXPERIMENTAL DETAILS}
Using the DUNE-ND neutrino(anti-neutrino) flux \cite{flux} in the energy range 0.125-20.125 GeV for Argon and Carbon nuclei, we have generated an inclusive data sample of 1 million charged current $\nu_{\mu}$ and $\bar\nu_{\mu}$ events. The analysis is performed by selecting pion-less events. The $\nu(\bar\nu)$ flux used in our work is shown in Figure 1, it covers the energy spectrum from hundreds of MeV to tens of GeV and peaks around 2.5 GeV decreasing significantly after that and becoming negligible after 10 GeV. The flux of neutrinos is slightly higher than the anti-neutrino flux around the peak energy. The primary beam of protons in the energy range 60-120 GeV coming out from the main injector accelerator is made to smash on a graphite target. This collision will result in the production of pions and kaons which will be further focussed with the help of magnetic horns toward a 200m long decay pipe where they will decay into neutrinos and leptons of all the flavors. In this way, the NuMI(Neutrinos at Main Injector) beamline facility at Fermilab will generate an intense, wide-band, high purity (anti)neutrino beam with an initial beam power of 1.2 MW. At 1.2 MW of beam power, 1.1 $\times$ $10^{21}$ protons are expected per year from the accelerator \cite{dune2020}. This facility will be further upgraded to 2.4 MW. The interaction cross sections($\nu_{\mu}(\bar\nu_{\mu}$)-Argon and $\nu_{\mu}(\bar\nu_{\mu}$)-Carbon) used in this work include Quasi-Elastic(QE), Resonance(RES)($\Delta$(1232) resonance production only and not higher mass resonances), two particle-two hole(2p2h/MEC) and Deep Inelastic Scattering(DIS) interaction processes in both neutrino and anti-neutrino modes. We have selected two different physics models to study the QE interaction process viz. (1) Llewellyn-Smith model (2) Nieves model.

By default in GENIE, the QE scattering is modeled using the Llewellyn-Smith model \cite{lewsmith} mentioned as the Default model in our work. While simulating the QE interactions the nuclear structure is described by the Relativistic Fermi Gas model(RFG) \cite{rfg1, rfg2}. Further we tune GENIE's description to simulate the QE interactions with the Nieves model \cite{nieves2} which is based on RPA effect. The Nieves model  describes 0 pion events in such a way that QE(1p-1h) cross-section gets modified in a non-trivial way. The movement of nucleons in the nuclear environment is treated differently by the Default model and the Nieves model. The models differ in actual distributions of nucleon momenta. The nucleus is treated according to Global Fermi Gas(GFG) by the Default model while the Nieves model treats the nucleus using a Local Fermi Gas(LFG) distribution. The GFG model implements a sphere of radius 250 MeV with momentum uniformly distributed inside the sphere. This approach gives the simplest momenta distribution. In the LFG model, a varying local nuclear density is seen by different nucleons and it depends on the distance of the nucleon from the center of the nucleus. Nuclei with higher density have a higher fermi limit. Momenta distribution predicted from both the models can be found in \cite{juan}.

\begin{figure}
 \centering\includegraphics[scale=.44]{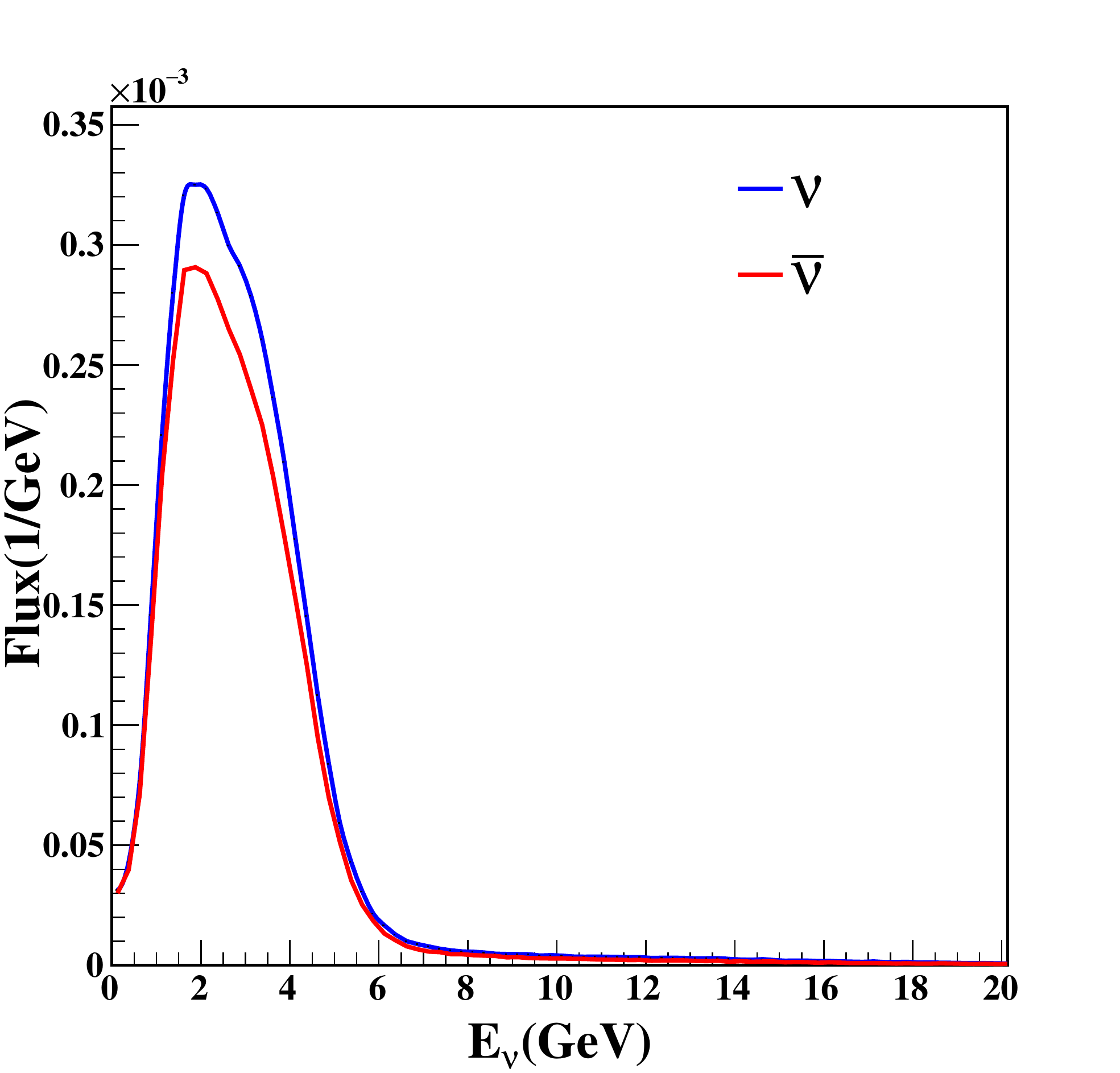}
 \caption{The DUNE flux as a function of neutrino(blue line) and anti-neutrino(red line) energy used in our work.}
\end{figure}

The vector form factor in GENIE is defined using BBBA05 \cite{bbba05} and the value of axial mass $M_{A}$ considered in this work is 0.99 GeV/$c^{2}$. For modeling 2p2h interactions, GENIE implements the model proposed by the Valencia group \cite{2p2hmodel}. The Rein-Sehgal model \cite{reinsehgal} is used for modeling the production of baryonic resonances in the neutral and charged current interaction channels. For the modeling of FSI i.e. simulating re-scattering of pions and nucleons in the nucleus, GENIE uses hA and hN as FSI models or the simulation package INTRANUKE \cite{intranuke,intranuke2}. The DIS interaction is simulated using the Bodek and Yang model\cite{dis}.

\begin{figure}
 \centering\includegraphics[width=8.0cm,height=7.0cm]{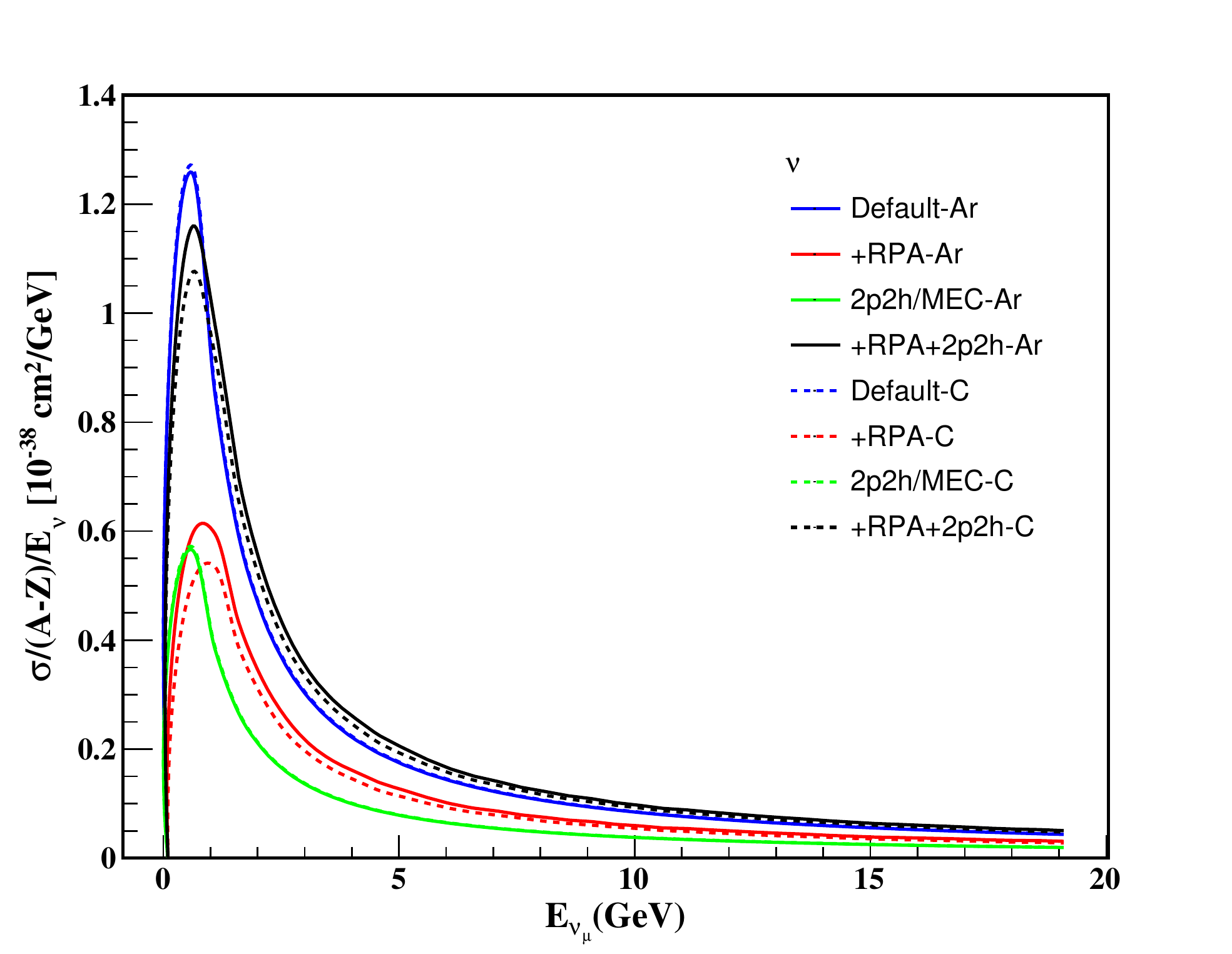}
\centering\includegraphics[width=8.0cm,height=7.0cm]{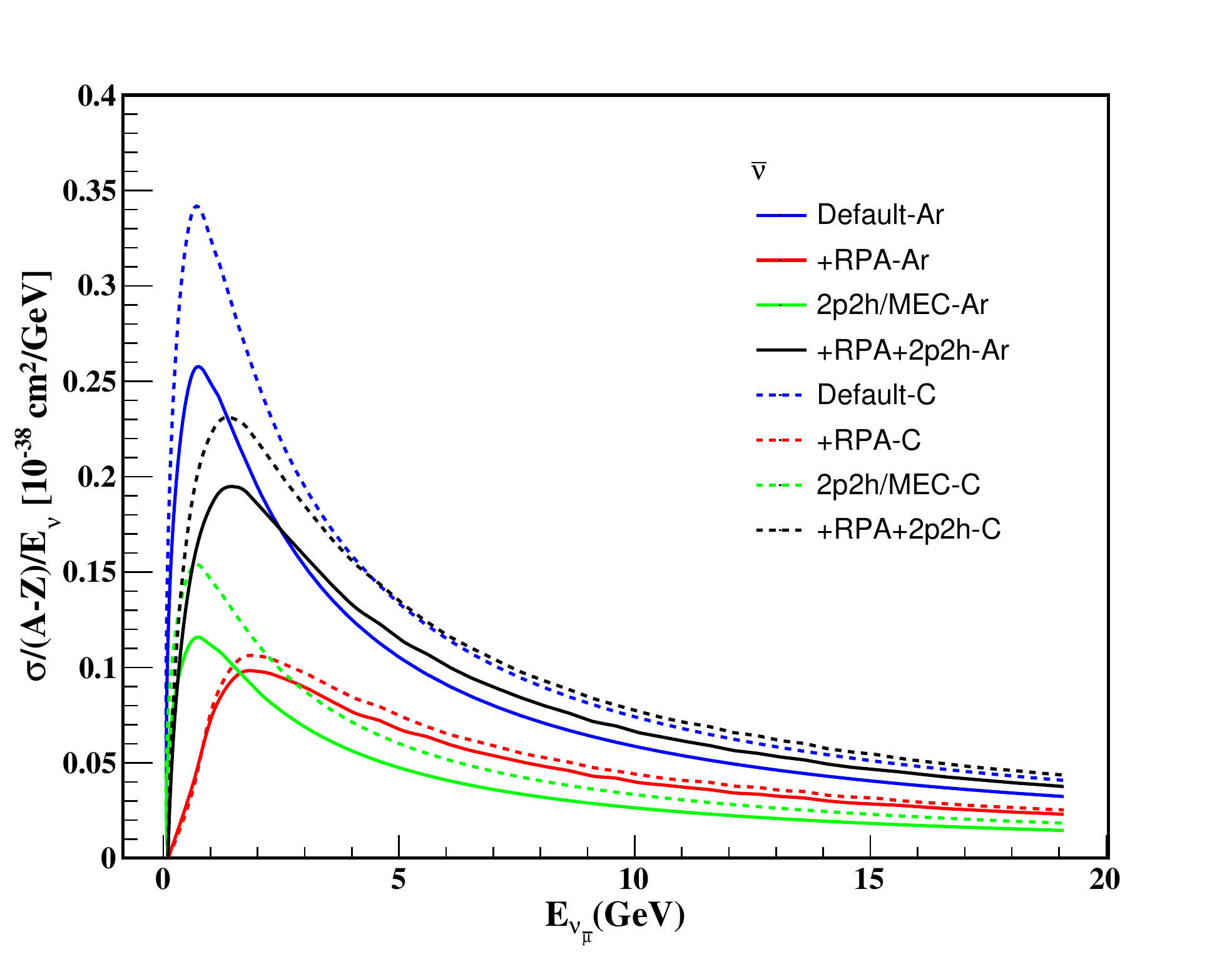}
\caption{The $\nu$-Argon(solid lines) and $\nu$-Carbon(dashed lines) integrated cross-section for the Default model(blue line), the same with RPA suppression(red line), with only 2p2h model(green line) and for a RPA combined with 2p2h component(black line) in the left panel and the same for $\bar\nu$-Argon(solid lines) and $\bar\nu$-Carbon(dashed lines) in the right panel.}
\end{figure}

The variation in $\nu_{\mu}(\bar\nu_{\mu})$-Argon(solid lines) and $\nu_{\mu}(\bar\nu_{\mu})$-Carbon(dashed lines) integrated cross section generated using different models as a function of neutrino energy is shown in Figure-2. The left panel is for neutrinos and the right panel is for anti-neutrinos. In Figure 2, we can observe the difference in the cross-section prediction for the QE process induced by the weak charged current interaction of $\nu_{\mu}(\bar\nu_{\mu})$ scattering on Argon nuclei relevant for DUNE. We can notice that neutrino cross-section estimated using the Default model for both the targets is almost the same but with the Nieves model the cross-section for Carbon is slightly lesser as compared to Argon. 

The difference in the cross-sections(Default model, model with RPA effect, 2p2h and RPA+2p2h interactions) for $\nu_{\mu}(\bar\nu_{\mu})$-Argon and $\nu_{\mu}(\bar\nu_{\mu})$-Carbon interaction at low $\nu(\bar\nu)$ energies i.e. for energies $\lesssim$ 1GeV is prominent. From Figure 2, we can observe that on combining the cross section obtained from the model including the RPA effect(represented by red line) with the cross-section contribution arising from the 2p2h process(represented by green line) shown by a black line, is comparable with the CCQE cross-section as obtained from the Default model in GENIE(represented by blue line) for neutrinos, but the same combination for anti-neutrinos shows some difference. From the right panel of Figure 2, we observe that the $\bar\nu$-Carbon cross-section is higher than the $\bar\nu$-Argon cross-section when compared with the Default, RPA, 2p2h and RPA+2p2h models.

The simulated final state particles are categorized in two ways- (i)When no detector cuts are imposed, the particles are represented as "Calculated" or "Cal" (ii)When detector cuts are imposed, the particles are represented as "Visible" or "Vis". Since not all the particles that emerge out of the various interaction processes can be detected by the detector, in order to find out the visible particles or those particles that can be seen by the detector, we apply certain kinetic energy thresholds on the particles which are mentioned in Section-II.

The kinematics of the interactions is described as follows: $E_{\nu}$ = $E_{\mu}$ + $q_{0}$ where $E_{\nu}$ is the energy of the neutrino, $E_{\mu}$ is the energy of the muon and $q_{0}$ is the energy transfer. The squared four momentum transfer $Q^{2}$ is calculated as- $Q^{2}$ = 2$E_{\nu}(E_{\mu}-p_{\mu}cos \theta_{\mu}) - M^{2}_{\mu}$ where $M_{\mu}$, $p_{\mu}$, $E_{\mu}$ and $\theta_{\mu}$ are the mass, momentum, energy and angle of the outgoing muon. The three momentum transfer is calculated as $q_{3} = \sqrt{Q^{2} + q_{0}^{2}}$.

\section{Multi-nucleon Interactions and their Implications}

A multi-nucleon interaction involves the interaction of a neutrino with multiple nucleons at the primary neutrino-nucleon interaction vertex. If an interaction involves only 1 nucleon then it is referred to as 1p1h(1particle-1hole) process, if 2 nucleons are involved then it is referred to as 2p-2h process or meson exchange currents(MEC). Further generalization of such a process to 'n' nucleons is referred to as 'np-nh' process. A significant contribution in the number of events arising from multinucleon interaction to the inclusive neutrino charge current cross-section is studied in \cite{multin4}, particularly for energies up to 1 GeV. The CCQE interactions occurring in this energy region(up to 1 GeV) are contaminated with multi-nucleon final states giving rise to the CCQE-like events. CCQE-like events are those events that at the initial state of interaction are not pure QE process but the final state products are similar to that of the pure QE interaction. For instance, in the case when a neutrino interacts with a correlated pair of nucleons, in the event of two nucleons being ejected out of the nucleus, such interaction will be classified as a 2p2h interaction. But in case if the second nucleon is absorbed within the nuclear environment due to FSI, or if it is a neutron or if the detector is incapable of detecting it owing to its low energy, then a 2p2h interaction would be a fake 1p1h interaction.

\begin{figure}
  \centering\includegraphics[scale=.69]{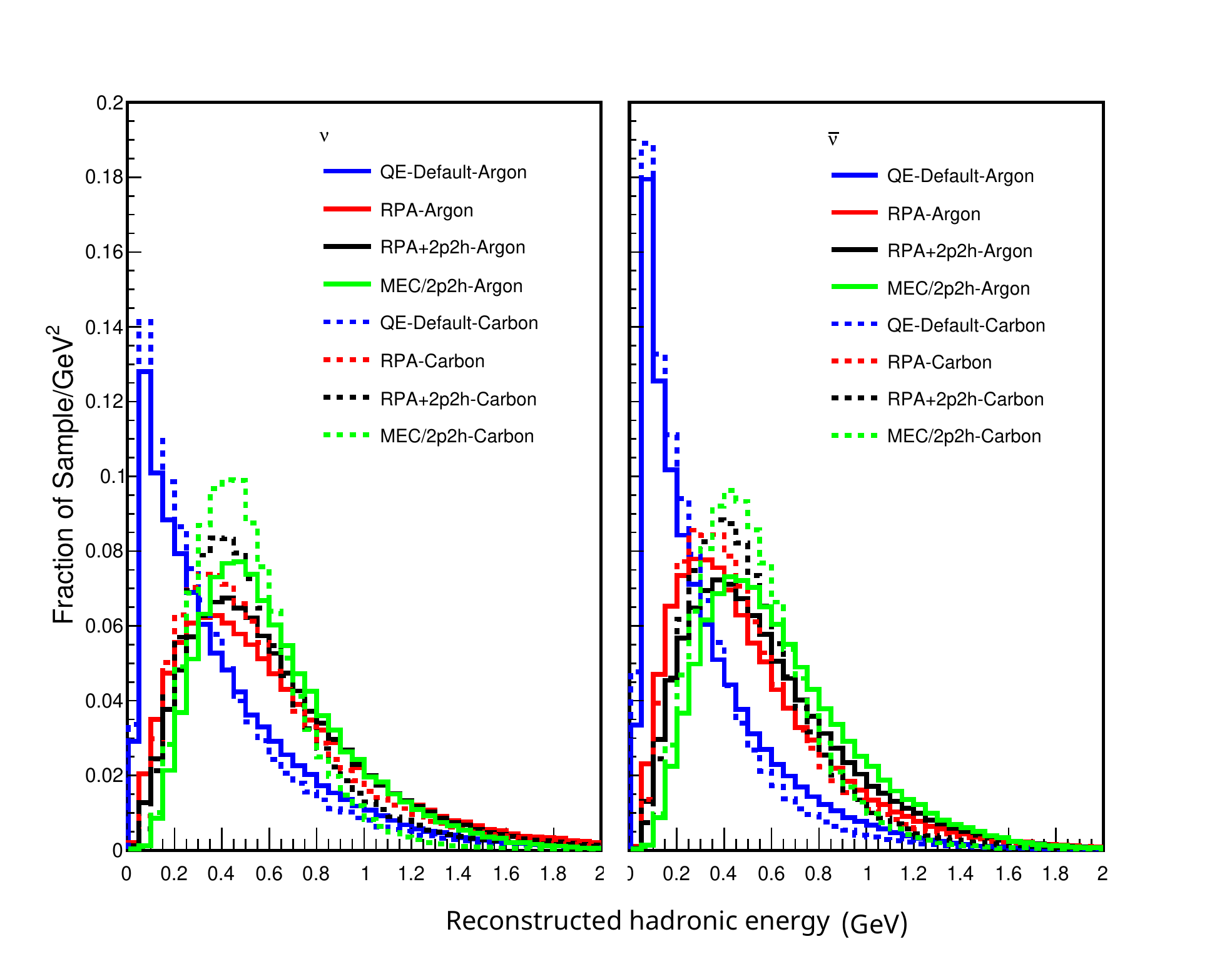}
  \caption{The left panel represents the reconstructed available hadronic energy($E_{had}$) for QE-Default(blue line), RPA(red line), MEC/2p2h(green line) and RPA+2p2h(black line) for $\nu$-Argon(solid lines) and $\nu$-Carbon(dashed lines). The right panel represents the same for $\bar\nu$-Argon and $\bar\nu$-Carbon}
\end{figure}
 
We know that the reconstruction of neutrino energy from the kinematics of the interaction products is of prime importance for precision oscillation analysis. We can employ either of the two methods for neutrino energy reconstruction i.e. (1) Kinematic method \cite{k1,k2,k3,k4} (2) Calorimetric method\cite{missenergy,calor}.
The kinematic method of energy reconstruction is based on lepton kinematics i.e. it requires information of the angle and momentum of the outgoing lepton. This method is based on the assumption that the incoming neutrino interacts with a single nucleon which is assumed to be at rest and the outgoing particles include a single nucleon and a lepton. As the energy of the incoming neutrino increases, contribution from other processes like RES, DIS come into existence and may lead to many hadrons in the final state thus leading to incorrect estimation of neutrino energy.\\
Contrary to the kinematic method, the calorimetric method of neutrino energy reconstruction requires information of all the visible final state particles on an event by event basis i.e. summing up of the energy deposited by all final state particles in the detector and thus allowing for accurate reconstruction of neutrino energy. It is applicable to all types of final states unlike the kinematic method which can be used only for QE event reconstruction. But even this method poses challenges in the way of precise neutrino energy reconstruction. One of the challenges is the accurate reconstruction of the final state hadrons, where neutrons and very low energy protons might escape detection. The calorimetric method thus depends upon the capability of the detector in reconstructing hadronic final states. The reconstruction of neutrino energy is also hindered due to the inevitable nuclear effects that may lead to a non-negligible amount of missing energy \cite{missenergy}

Applying the calorimetric approach i.e. summing up all the outgoing particles, reconstructed neutrino energy, $E_{\nu}^{Calor}$ \cite{missenergy} can be calculated as-\\
\begin{equation}
 E_{\nu}^{Calor} = E_{lep}+ \sum \limits_{i} T_{i}^{nuc} + \epsilon_{nuc} + \sum \limits_{m} E_{m}
\end{equation}
where $E_{lep}$ is the outgoing final state charged lepton's energy, $T_{i}^{nuc}$ is the kinetic energies of the outgoing nucleons(i.e. the protons and/or neutrons), their corresponding separation energies represented as $\epsilon_{nuc}$ and total energy of any other particle produced represented as $E_{m}$.
We can also write Equation(1) as- $E_{\nu}^{Calor} = E_{lep}+E_{had}$, where,
\begin{equation}
 E_{had} = \sum \limits_{i} T_{i}^{nuc} + \epsilon_{nuc} + \sum \limits_{m} E_{m}
\end{equation}
The distribution of reconstructed available hadronic energy($E_{had}$) in the detector is calculated using equation(2). The $E_{Had}$ is the energy accumulated by the hadrons in the detector and it is calculated by summing up kinetic energies of proton, charged pion and neutral pion. Figure 3 shows the reconstructed available hadronic energy calculated from different models for Argon(solid lines) and Carbon(dashed lines) using the Calorimetric technique. The left and the right panels are for neutrinos and anti-neutrinos respectively.

\section{RESULT AND DISCUSSION}
From Figure 3 we note that on considering the RPA effect by tuning the Nieves model in our simulation we observe a significant shift in the hadronic energy distribution pattern towards the higher energy, roughly by $\approx$ 300 MeV from the hadronic energy distribution pattern obtained from the Default model. In Figure 4 and Figure 5, where total event distribution is plotted as a function of $Q^{2}$ and $q_{3}$ respectively, we can observe a similar shift in the event distribution pattern, since these kinematics variables directly depend on hadronic energy.

It is evident from Figures 4 and 5, on implementation of the model with RPA effect(green and black lines) there is a difference in the total event distribution pattern as a function of $Q^{2}$ and $q_{3}$ as compared to the event rate distribution pattern that results from the Default model(blue and red lines) for $\nu(\bar\nu)$-Argon and $\nu(\bar\nu)$-Carbon. Since RPA correlations describe strongly interacting nucleons, these are responsible for changing the electroweak coupling strengths of the bound nucleons with respect to the free nucleon values. RPA modifies the low lying states of the nucleus by changing the potential which causes a screening effect. This screening effect reduces the probability of the weak interaction which results in a reduction of QE weak interaction cross-section. We can also notice this trend from Figure 2(left and right panels), which depicts the reduced cross-section on the implementation of the RPA model(red line) when compared to the Default model(blue line) for both neutrino and anti-neutrino. Therefore in the above $Q^{2}$ distribution plots, we can observe a reduction in the event rates when RPA effect is taken into account. We also note from the Figures 4 and 5, that $\bar\nu$-Argon and $\bar\nu$-Carbon event rates as a function of $Q^{2}$ and $q_{3}$ are higher than the $\nu$-Argon and $\nu$-Carbon event rates. This enhancement is significant when the Default model is implemented while with RPA correlations the enhancement is marginal for both Argon and Carbon. The difference between neutrino and anti-neutrino can be attributed to the fact that neutrons give different contributions to the final state energy. One can also study the energy spectrum of individual hadrons(neutrons and protons) as a function of neutrino energy in the reference \cite{calor}.

\begin{figure}
  \centering\includegraphics[width=8.0cm,height=7.0cm]{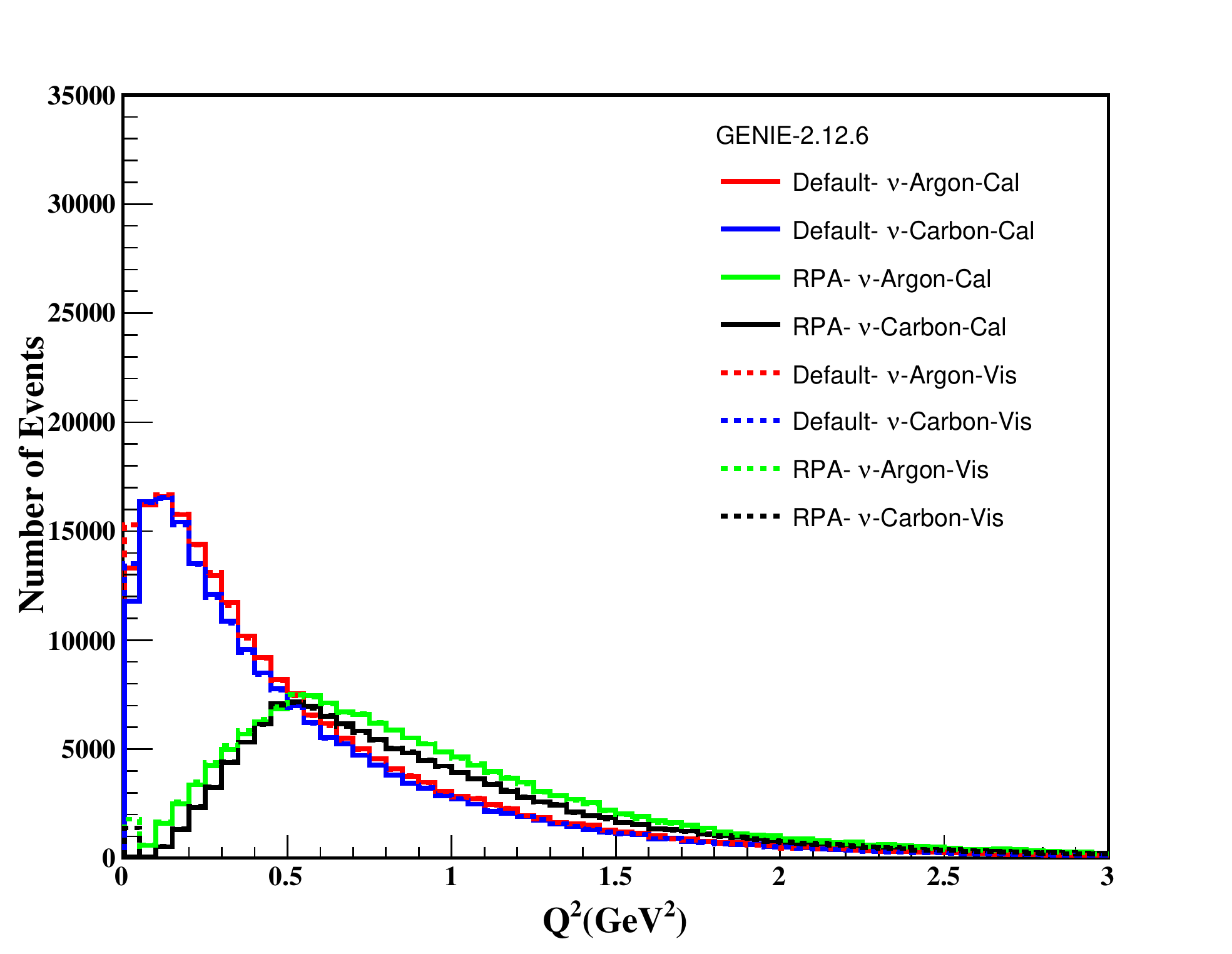}
   \centering\includegraphics[width=8.0cm,height=7.0cm]{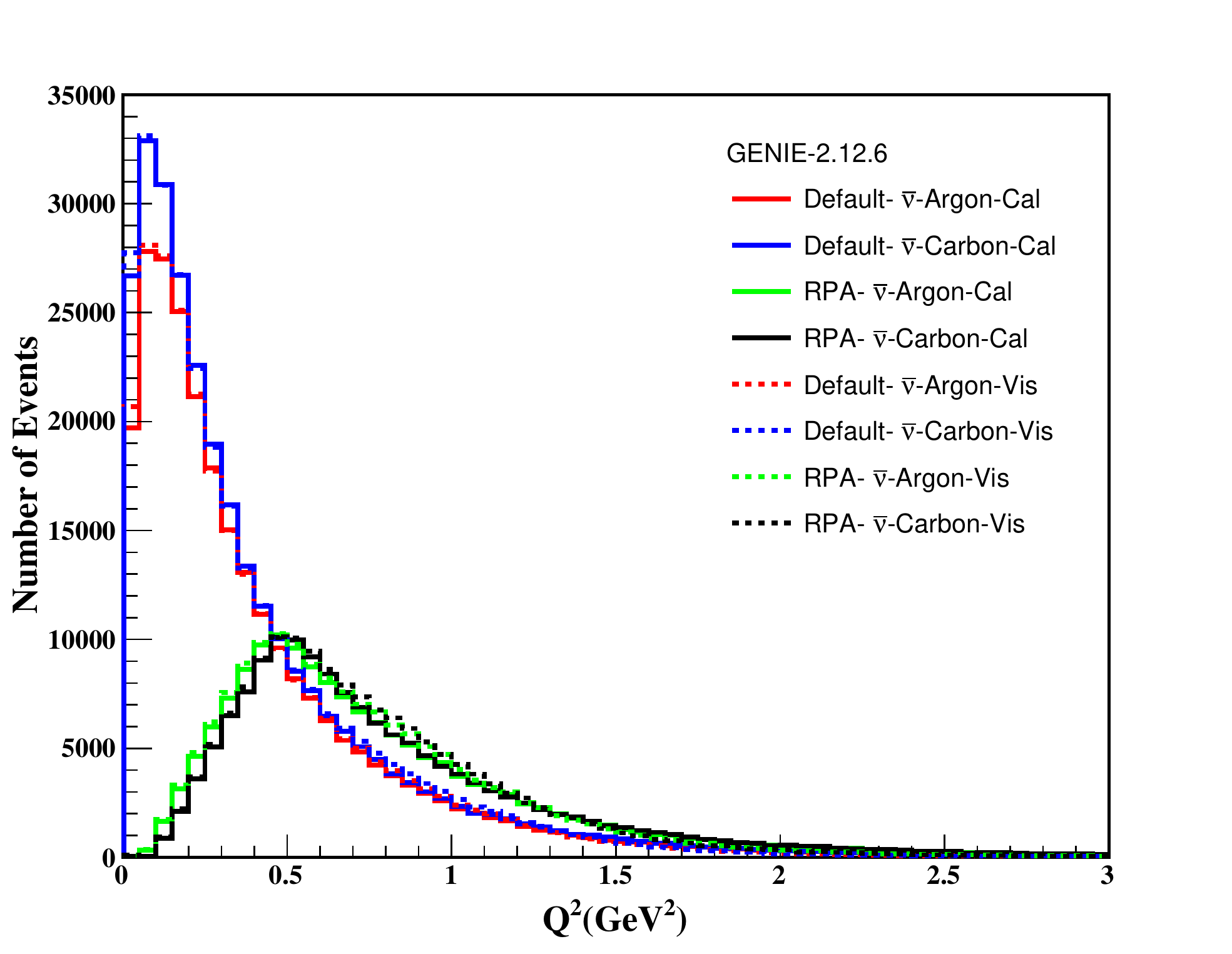}
  \caption{The left panel shows the event distribution as a function of squared four momentum transfer with the Default model and the RPA model for $\nu$-Argon and $\nu$-Carbon interactions while the same is shown for $\bar\nu$-Argon and $\bar\nu$-Carbon in the right panel. Note that "Cal" and "Vis" distributions largely overlap.}
\end{figure}

\begin{figure}
  \centering\includegraphics[width=8.0cm,height=7.0cm]{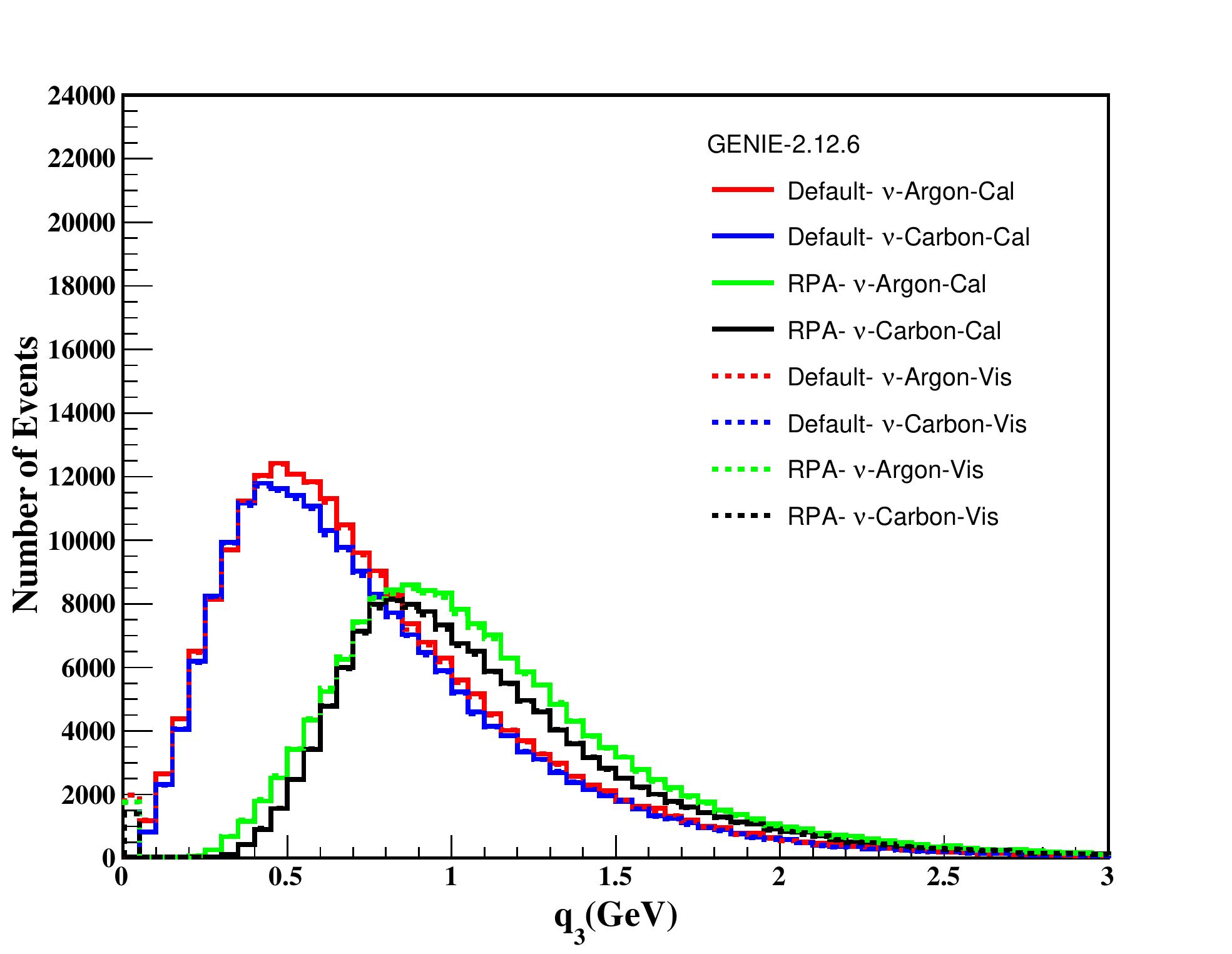}
  \centering\includegraphics[width=8.0cm,height=7.0cm]{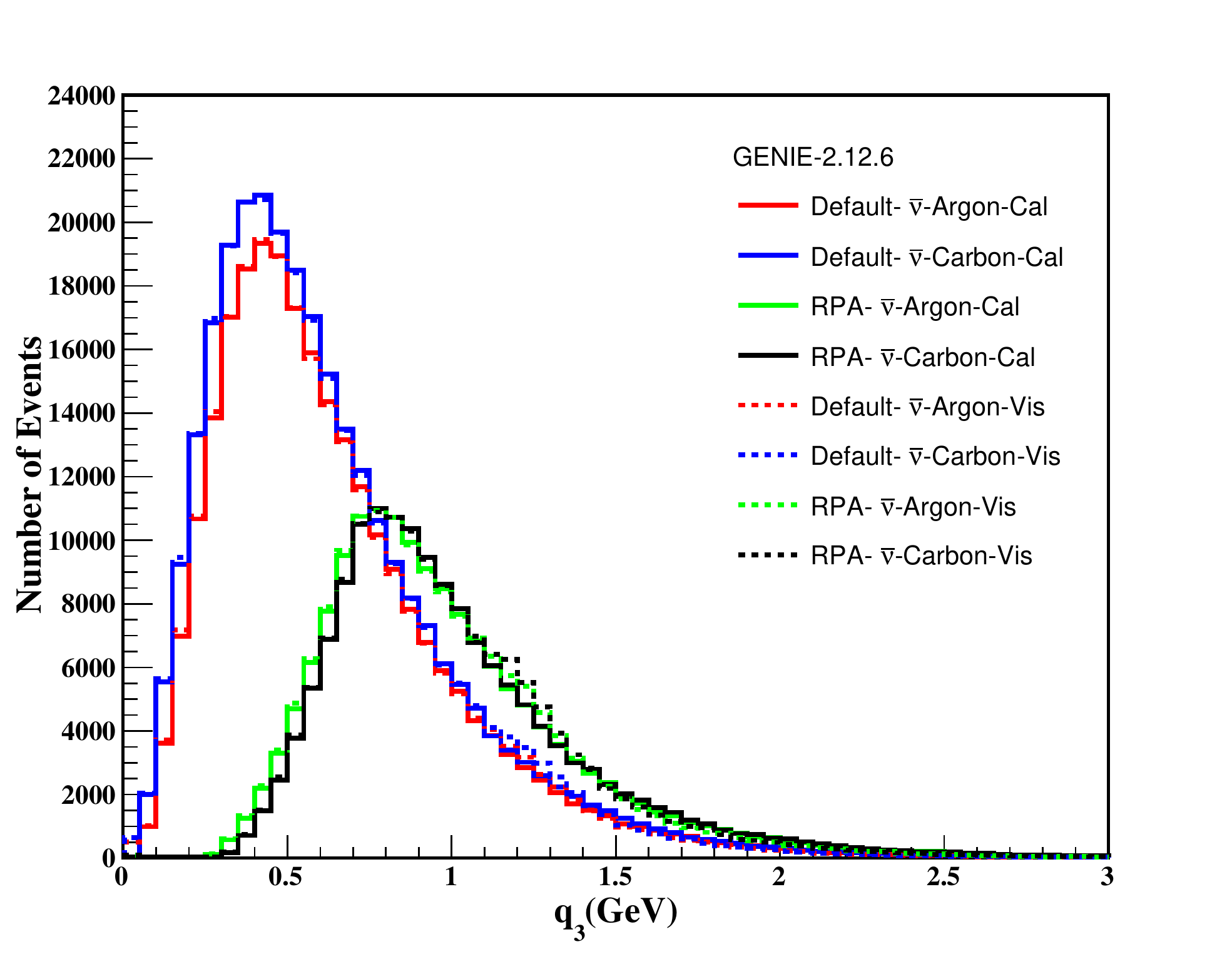}
  \caption{The left panel shows the event distribution as a function of three momentum transfer with the Default model and the RPA model for $\nu$-Argon and $\nu$-Carbon interactions while the same is shown for $\bar\nu$-Argon and $\bar\nu$-Carbon in the right panel. Note that "Cal" and "Vis" distributions largely overlap.}
\end{figure}
 
In an attempt to quantify the systematic uncertainties introduced due to nuclear effects we have estimated the ratio of events as a function of $Q^{2}$ for the two targets viz., Argon and Carbon and presented in Figure 6. The ratio of Ar/C is calculated to address the difference that will incur in the physics results due to the use of different target material. The ratio of events as a function of $Q^{2}$ for $\nu$-Ar/C is shown in the left panel of Figure 6, whereas the right panel shows the distribution of $\bar\nu$-Ar/C. We have checked the difference arising due to the consideration of the same models in different target materials. In the left panel, the blue lines(solid-Cal and dashed-Vis) represent the ratio of $\nu$-Ar/C for the Default model whereas the red lines(solid-Cal and dashed-Vis) represent the ratio of  $\nu$-Ar/C for the RPA model, the right panel has the same description for anti-neutrinos. We notice that the model with RPA effect shows minimal distortion from unity ($\leq$ 1 $GeV^{2}$) for both Cal and Vis datasets. But the Default model shows much deviation from unity and a dip around 0.5 $GeV^{2}$ indicating larger systematic uncertainty. The event distribution ratio as a function of $Q^{2}$ presented in the right panel for anti-neutrinos shows some symmtery with the RPA model below 1 $GeV^{2}$.

\begin{figure}
  \centering\includegraphics[width=8.0cm,height=7.0cm]{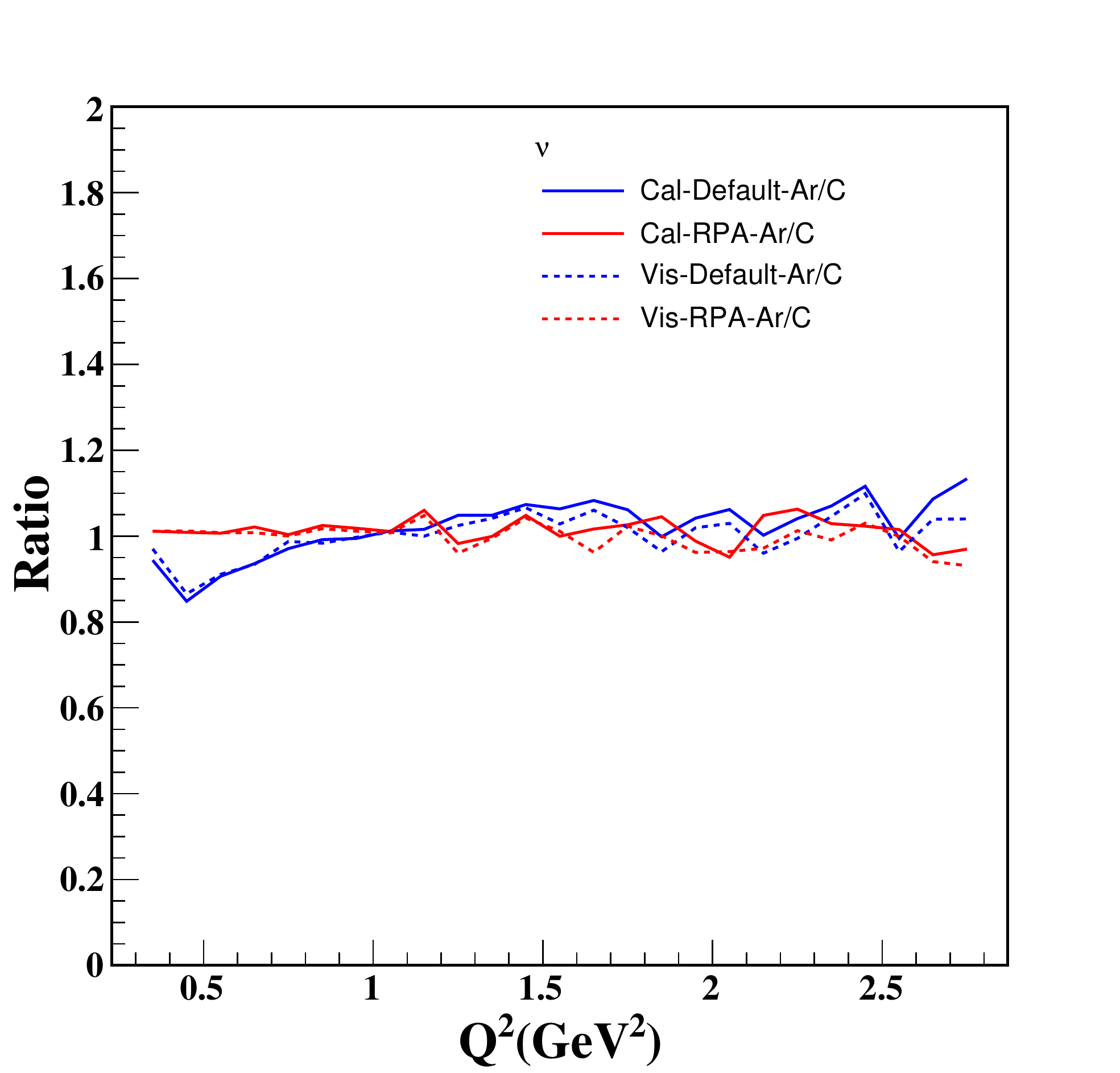}
  \centering\includegraphics[width=8.0cm,height=7.0cm]{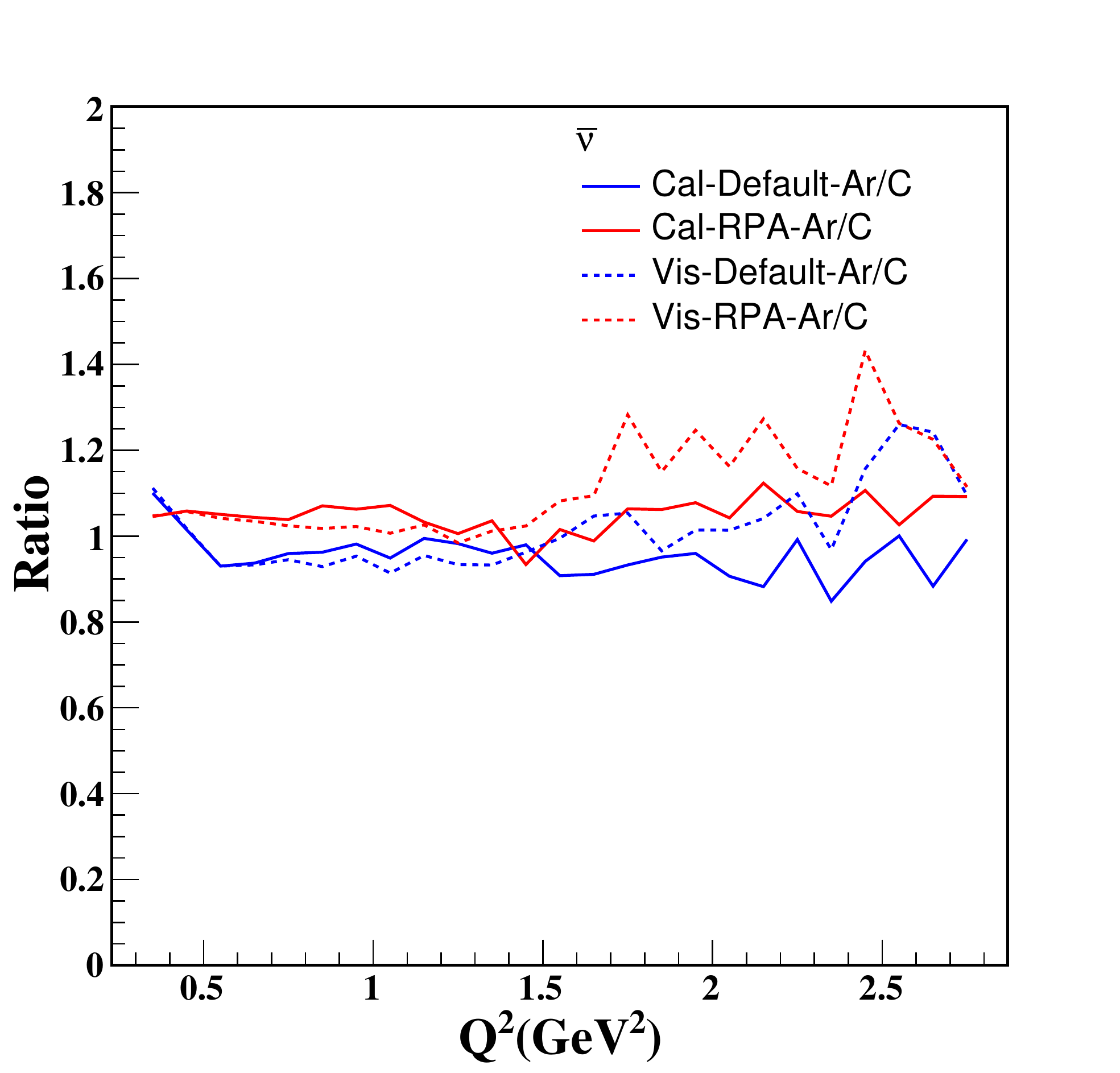}
  \caption{The left panel shows the ratio of $Q^{2}$ distribution for Ar/C achieved using the Default model and model with RPA effect for neutrinos and the same is displayed in the right panel for anti-neutrinos. Solid lines represent 'Calculated' particles while the dashed lines represent the 'Visible' particles.}
\end{figure}

\begin{figure}
  \centering\includegraphics[width=8.0cm,height=7.0cm]{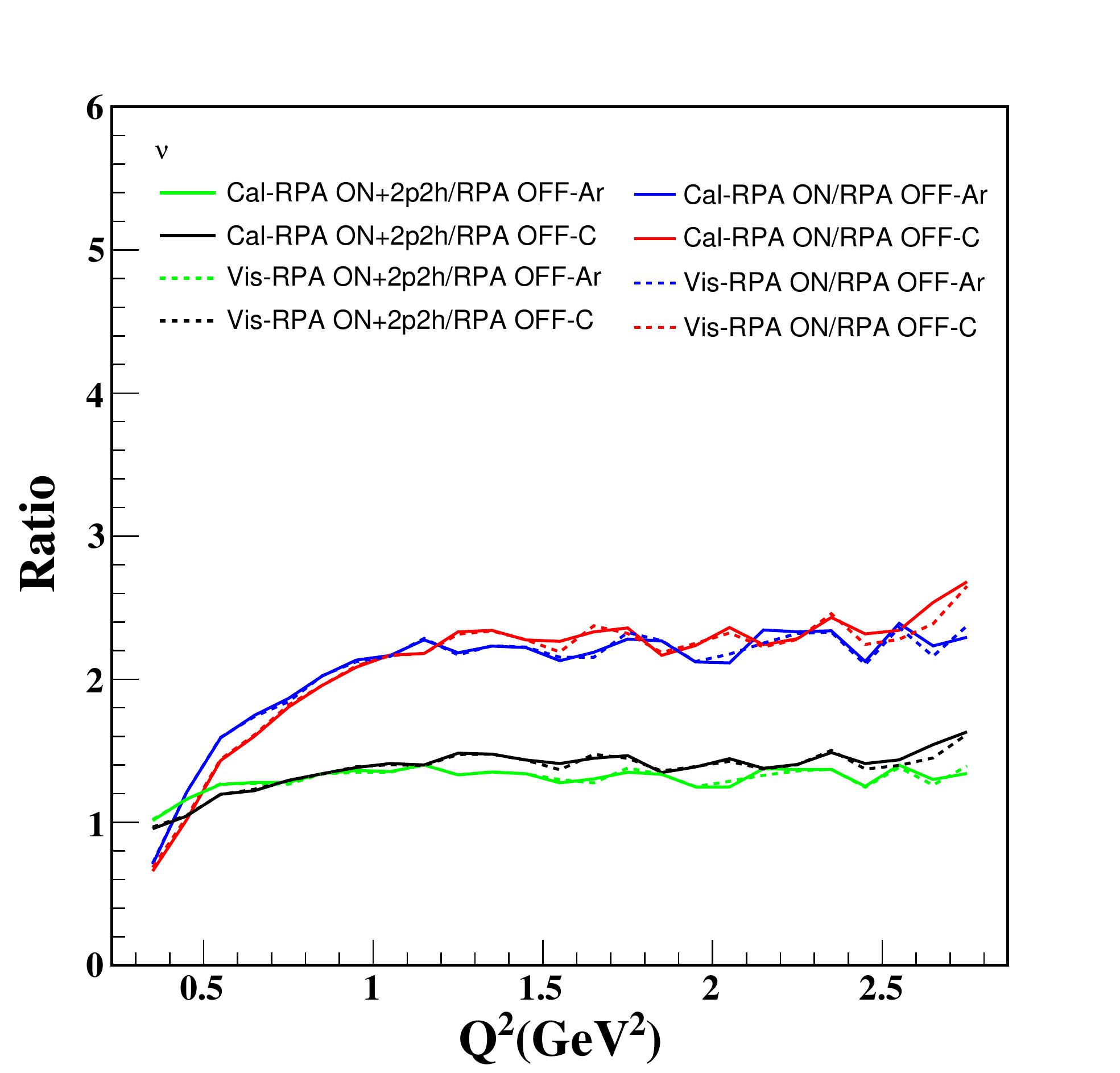}
  \centering\includegraphics[width=8.0cm,height=7.0cm]{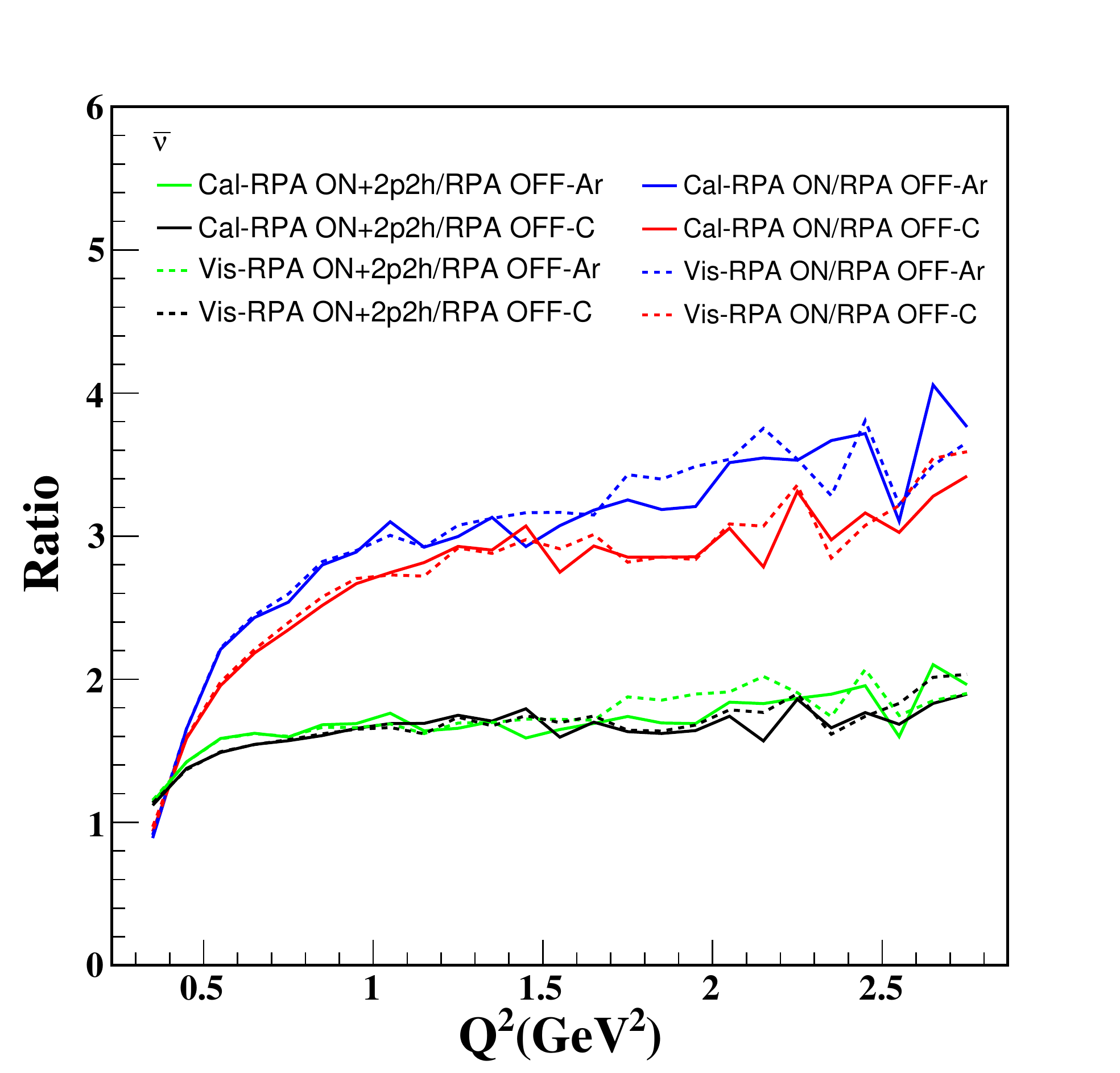}
  \caption{The left panel shows the ratio of $Q^{2}$ distribution for $\nu$-Ar/Ar(blue line) and $\nu$-C/C(red line) with RPA effect ON and RPA effect OFF and the ratio of $Q^{2}$ distribution for RPA ON+2p2h and RPA OFF for Ar/Ar(green line) and C/C(black line), the same is represented for anti-neutrinos in the right panel. Solid lines represent 'Calculated' particles while the dashed lines represent the 'Visible' particles.}
\end{figure}

Further to check the systematic uncertainty in the models itself, we have estimated the ratio of event distributions as a function of $Q^{2}$ with RPA effect ON(RPA-ON) to RPA effect OFF (RPA-OFF) for the same target material i.e. Argon and Carbon and is represented in the left($\nu$) and right($\bar\nu$) panels of Figure 7. We found that the ratio of RPA-ON/RPA-OFF(blue and red lines for Argon and Carbon respectively) as a function of $Q^{2}$ for the same target increases continuously from 0 to 1 $GeV^{2}$ which implies a significant amount of uncertainties below 1$GeV^{2}$. 
Whereas on the addition of 2p2h contribution(indicated by green and black lines for Argon and Carbon respectively) to the MC datasets(both Cal and Vis), a significant improvement in the model prediction is observed. We notice that the ratio of RPA-ON+2p2h/RPA-OFF becomes nearly constant for values between 1.2 to 1.4 but for anti-neutrinos this range falls between 1.4 to 1.8. Also for neutrinos, the Cal and Vis data sets mostly overlap but a discrepancy is observed for anti-neutrinos where Vis and Cal datasets have more distortions.

\section{Conclusions}
There is a dearth of knowledge about the interaction of neutrinos with nucleons bound inside the nucleus. To understand neutrino-nucleon interactions, we should have a good understanding of the hadronic physics of these interactions since it is the key to understand nuclear effects. The challenge is to develop the models which can describe the complete time-development of a neutrino-nucleon interaction for all the processes. There are large uncertainties in $(\bar\nu)\nu$-nucleon cross section in the scattering region governed by the QE interactions. Reaction mechanisms involving two particles and two holes that arise in the target nucleus are one of the sources of systematic uncertainties in the QE scattering region and are required to be addressed carefully as they play a significant role in the energy region relevant for the DUNE experiment \cite{duneNDref20}. Processes like multinucleon interactions are still neither modeled nor constrained properly in the event generators currently in use. We must attempt to tune the available models as much as possible so that the DUNE-ND complex can deliver precision physics.

We find from Figure 7, that the Nieves model along with the 2p2h interaction sample result in a marked reduction in systematic uncertainties for both neutrinos and anti-neutrinos. Thus, we can say that incorporation of 2p2h interactions combined with the RPA correlations may be considered by DUNE to reduce systematic uncertainties occurring in the low energy regime and at the same time, the models must be tuned to reduce the uncertainties and checked for both neutrinos and anti-neutrinos.

\section{ACKNOWLEDGEMENT} 
This work is supported by the Department of Physics, University of Lucknow. Financially it is supported by the Government of India, DST Project no-SR/MF/PS02/2013, Department of Physics, University of Lucknow.

\end{document}